\pdfoutput=1
\documentclass[conference]{IEEEtran}
\IEEEoverridecommandlockouts

\usepackage{cite}
\usepackage{amsmath,amssymb,amsfonts}
\usepackage{algorithmic}
\usepackage{graphicx}
\usepackage{textcomp}
\usepackage{xcolor}
\usepackage{booktabs}
\usepackage{multirow}
\usepackage{url}

\begin{document}

\title{Graph Attention Design Choices Matter: A Controlled Study of LoRA-Adapted Audio Anti-Spoofing}

\author{
\IEEEauthorblockN{Haoyu Wang\textsuperscript{1,2}, Jing Yang\textsuperscript{2,3}, Chenyu Liu\textsuperscript{2}, Yushan Du\textsuperscript{1}, Yifan Liao\textsuperscript{2,4},\\
Ningning Pan\textsuperscript{1,*}, Gongping Huang\textsuperscript{3}, Yu Zhao\textsuperscript{1}, Gang Li\textsuperscript{2}, and Jian Luan\textsuperscript{2}}
\IEEEauthorblockA{\textsuperscript{1}School of Computing and Artificial Intelligence,\\
Southwestern University of Finance and Economics, Chengdu, China\\
\textsuperscript{2}MiLM Plus, Xiaomi Inc., Beijing, China\\
\textsuperscript{3}School of Electronic Information, Wuhan University, Wuhan, China\\
\textsuperscript{4}NERCMS, School of Computer Science, Hubei Luojia Laboratory,\\
Wuhan University, Wuhan, China}
\thanks{\textsuperscript{*} Corresponding author.}
}

\maketitle
\begin{abstract}
Audio anti-spoofing systems increasingly combine self-supervised learning, parameter-efficient fine-tuning, and graph-attention-based backends. However, performance gains in such systems are often entangled with concurrent changes in the backbone, fine-tuning strategy, and training protocol, making the independent contribution of graph attention design difficult to isolate. To address this issue, we conduct a systematic controlled study of the graph attention layer under a unified experimental setting. We decompose the layer into three independently testable design dimensions: scoring symmetry, temperature learnability, and routing granularity. These are instantiated as a concat-based scoring branch, a LearnT branch with learnable temperature, and a multi-temperature routing branch, respectively. Each dimension is implemented as an independently gated residual branch, enabling the evaluation of both individual variants and their combinations under the same experimental setting. Experiments on five evaluation sets with five random seeds show that the LearnT branch achieves the best average equal error rate (EER), yielding a $16.1$\% relative improvement over the baseline. In contrast, the multi-temperature routing branch does not improve average performance on its own, but substantially reduces cross-seed standard deviation when combined with the concat-based scoring branch. Moreover, two individually effective branches degrade performance when used together, resulting in a $25.6$\% relative deterioration compared with the baseline. This finding reveals strong non-additive interactions among graph attention design dimensions. Overall, the results suggest that, under parameter-constrained fine-tuning, improvements in graph attention layers depend more on capacity allocation and branch interaction than on simply adding more learnable parameters.

\end{abstract}


\section{Introduction}

\begin{figure}[!t]
  \centering
  \includegraphics[width=\columnwidth]{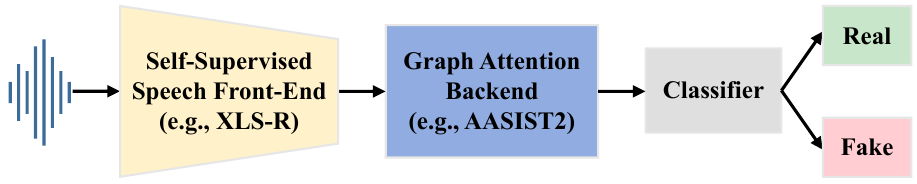}
  \caption{Overview of the LoRA-adapted audio anti-spoofing framework used in this study.}
  \label{fig:overview_intro}
\end{figure}

Recent advances in generative audio have made synthetic speech increasingly difficult to distinguish from genuine recordings~\cite{li2025csur_speechdeepfake_survey}, thereby posing serious threats such as fraud and the dissemination of disinformation. In audio anti-spoofing, a common paradigm combines a self-supervised speech front-end~\cite{wang2022odyssey_ssl_frontends,tak2022odyssey_wav2vec2,zhang2024mm_xlsr_sls,guo2024icassp_wavlm_mfa}, such as XLS-R~\cite{babu2022interspeech_xlsr}, with a graph-attention-based backend, such as AASIST2~\cite{zhang2024icassp_aasist2}. Recent parameter-efficient systems further adapt SSL front-ends with LoRA-style modules~\cite{hu2022iclr_lora,laakkonen2025arxiv_metalora,xia2024taslp_molex,chen2025amulet}, while AASIST variants continue to modify the graph-attention backend~\cite{viakhirev2025arxiv_scalable,borodin2024asvspoof_aasist3}. In our controlled setting, LoRA modules are inserted into the Transformer layers of the front-end and trained together with LayerNorm parameters, while the original pretrained backbone weights remain frozen, reducing the number of trainable parameters to below $1$\% of the full model, as illustrated in Figure~\ref{fig:overview_intro}. Although these systems have further extended the overall SSL-plus-backend framework, the graph-attention backend has largely been treated as a fixed component, and its internal design choices have rarely been examined in a direct and systematic manner.


To characterize this gap more precisely, we revisit the graph attention layer in AASIST2. This layer involves three implicit design choices: (i) product-based symmetric scoring (i.e., $e_{ij} = e_{ji}$), (ii) a fixed softmax temperature, and (iii) a single global temperature shared across all nodes. Related work has studied adjacent mechanisms in separate contexts: GATv2~\cite{brody2022iclr_gatv2} modifies graph-attention scoring, mixture-of-experts methods introduce routing mechanisms for graph neural networks~\cite{chen2025arxiv_gnnmoe}, GATS~\cite{hsu2022neurips_gats} learns node-wise temperatures for GNN calibration, and AASIST3~\cite{borodin2024asvspoof_aasist3} evaluates fixed-temperature variants in anti-spoofing. However, these factors have not been isolated as controlled design dimensions within LoRA-adapted AASIST2 audio anti-spoofing.

To address this gap, we conduct a controlled factorization of the graph-attention design space under a unified experimental setting. Specifically, we decompose it into three independently testable dimensions: scoring symmetry, temperature learnability, and routing granularity, and instantiate each of them as a corresponding residual branch for systematic comparison. All variants are evaluated under the same adaptation protocol, enabling a more direct analysis of the effects of individual modifications and their combinations.

Systematic evaluations on one in-domain test set, four out-of-domain test sets, and five random seeds reveal pronounced non-additive interactions among these design dimensions. The LearnT branch achieves the best average EER. By contrast, the multi-temperature routing branch does not improve average performance when used in isolation, but yields the lowest cross-seed standard deviation when combined with the concat-based scoring branch. More notably, two modifications that are each effective when introduced separately lead to a substantial performance degradation when applied jointly. Overall, these findings indicate that the effects of different design choices within graph attention are not independent, and that a systematic analysis of their interactions is important for improving current audio anti-spoofing models.

We summarize our contributions as follows:
\begin{enumerate}
\item We present a controlled decomposition of graph attention design choices in LoRA-adapted audio anti-spoofing into independently testable dimensions.
\item We provide empirical evidence that, within a unified LoRA-adapted AASIST2 setting, architectural modifications that are beneficial in isolation can lead to performance degradation when combined.
\item We show that LearnT achieves the best mean EER, while concat-based scoring combined with multi-temperature routing yields the lowest cross-seed standard deviation.
\end{enumerate}

\section{Related Work}
\paragraph{Graph Attention Architectures for Audio Anti-Spoofing.}
Building on graph attention networks~\cite{velickovic2018iclr_gat} and
early GAT-based anti-spoofing systems~\cite{tak2021interspeech_gat_asv},
AASIST~\cite{jung2022icassp_aasist} introduced heterogeneous graph attention networks to
audio anti-spoofing, jointly modeling spectral and temporal graph branches
with product-based symmetric scoring and a fixed softmax temperature. AASIST2~\cite{zhang2024icassp_aasist2}
improved the AASIST architecture with Res2Net-style residual blocks and
additive-margin softmax, but the internal graph attention mechanism
remained largely inherited from AASIST. Subsequent variants explored modifications to
this design, but without isolating individual factors. RawGAT-ST~\cite{tak2021asvspoof_rawgatst}
replaced the entire GAT module with a raw waveform GAT encoder and
spectro-temporal graph attention, altering both input representation
and attention mechanism simultaneously. Raw waveform front ends and
augmentation have also been explored for anti-spoofing~\cite{tak2021icassp_rawnet2,tak2022icassp_rawboost}.
The scalable AASIST line of
work~\cite{viakhirev2025arxiv_scalable} substituted the GAT layers with multi-head attention, and
AASIST3~\cite{borodin2024asvspoof_aasist3} evaluated HS-GAL branches with several fixed
attention temperatures (e.g., $T=100$, $150$, and $200$), treating
temperature as a manually selected hyperparameter rather than a trainable
parameter. SpAArSIST~\cite{firc2026spaarsist} recently simplified the
HS-GAL stack-node attention, finding that temperature scaling did not
improve their sparsified variant; however, this work examined attention
simplification at the backend-pooling level rather than a controlled
factorization of the original AASIST2 graph attention layer. Across these
efforts, little prior work has decomposed that layer's design choices
into separately testable dimensions, and evaluated each under a common
controlled protocol. This gap motivates our controlled decomposition of
scoring symmetry, temperature learnability, and routing granularity as
independent residual additions to a fixed baseline.

\paragraph{Scoring, Temperature, and Routing in Graph Attention.}
Beyond the anti-spoofing domain, each of our three design axes has
an established literature in graph representation learning. The
element-wise-product scorer inherited by AASIST-style graph attention is
necessarily symmetric ($e_{ij}=e_{ji}$), whereas
GATv2~\cite{brody2022iclr_gatv2} showed that concat-based scoring with
LeakyReLU broke such static ranking behavior and enabled query-adaptive
attention weights, though this was demonstrated under full-training
regimes on citation networks. Separately,
temperature scaling is a standard calibration mechanism~\cite{guo2017icml_calibration},
and GATS~\cite{hsu2022neurips_gats} demonstrated that node-wise temperatures
could be learned for GNN prediction calibration. Recent GAT variants have
also explored learnable temperature as an explicit attention-sharpness
control~\cite{ma2026arxiv_gated_temp_gat}. In the multi-expert
direction, sparsely gated and switch-style expert routing~\cite{shazeer2017iclr_moe,fedus2022jmlr_switch}
motivated mixture-of-experts designs for graph neural
networks~\cite{chen2025arxiv_gnnmoe}, and multi-temperature attention mechanisms were
explored for sequence models, but per-node temperature routing within a
graph attention layer remained largely untested in any application domain.
More broadly, recent work established that learnable scalar
parameters could improve performance through their effect on optimization
dynamics rather than their final values: Agarwala et al.~\cite{agarwala2023temperature}
showed that the softmax temperature governed the transition from lazy to
feature-learning regimes during training, and Wang et al.~\cite{wang2026scalevectors}
demonstrated that scale vectors in LLMs were expressively redundant yet
indispensable, improving optimization through a self-amplifying
preconditioning effect. While each of these mechanisms was individually
motivated and validated in its original context, none has been
systematically evaluated in the specific setting where they interact:
frozen pretrained SSL weights with trainable LoRA adapters and limited
trainable capacity.

\paragraph{LoRA Adaptation of Self-Supervised Speech Front-Ends.}
Self-supervised speech models such as
Wav2Vec~2.0~\cite{Baevski2020wav2vec,tak2022odyssey_wav2vec2,lee2022interspeech_wav2vec_sasv},
XLS-R~\cite{babu2022interspeech_xlsr,zhang2024mm_xlsr_sls},
WavLM~\cite{chen2022jstsp_wavlm,guo2024icassp_wavlm_mfa},
and HuBERT~\cite{HsuBTLSM21HuBERT} have become strong front-ends for
spoofing and audio deepfake detection. Parameter-efficient fine-tuning
provides a general framework for adapting large pretrained models with a
small trainable parameter budget~\cite{hu2022iclr_lora,lialin2023arxiv_peft_survey},
and recent audio deepfake systems have adopted LoRA-style adapters for
cross-dataset generalization. Meta-learned LoRA~\cite{laakkonen2025arxiv_metalora} framed
generalization across attacks as a domain-shift problem and applied
MLDG-style meta-optimization to the LoRA weights. MoLEx~\cite{xia2024taslp_molex}
and AMULET~\cite{chen2025amulet} similarly explored adapter placement,
expert fusion, and training strategies for cross-dataset generalization.
Among these works, the downstream classifier, often a graph attention
network, was treated as a fixed inheritance: LoRA rank, placement, and
training objectives were systematically explored, while the internal
scoring, temperature, and routing of the graph attention layers were
carried over from the baseline without question. We address this gap
by holding the LoRA configuration constant and systematically varying the
graph attention internals to identify which design choices are
consequential under the resulting capacity budget.

\section{Methodology}
\label{sec:methodology}

\begin{figure}[t]
\centering
\includegraphics[width=\columnwidth]{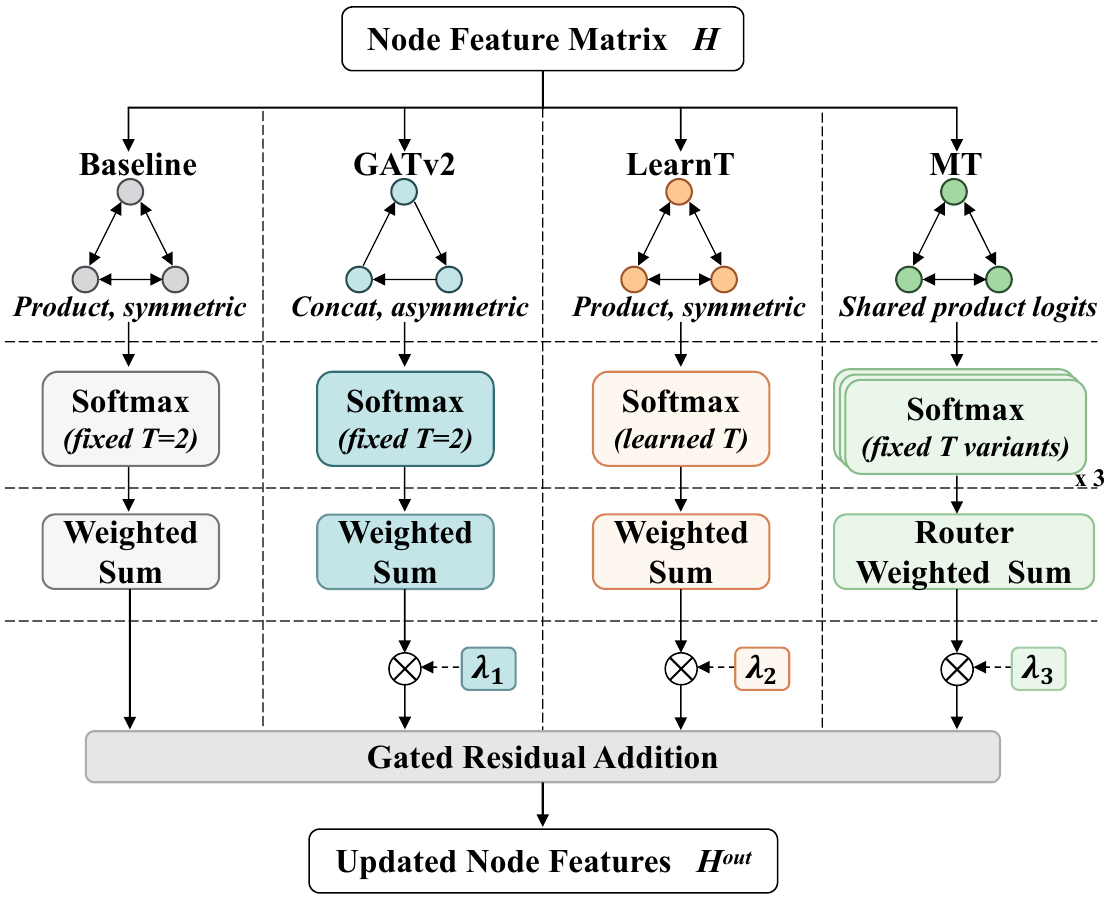}
\caption{Residual graph-attention framework for one AASIST2 graph-attention
layer. The baseline uses symmetric element-wise-product scoring with fixed
temperature $T_0=2.0$, while three gated residual branches instantiate the
design axes: GATv2 for concat-based asymmetric scoring, LearnT for learnable
softmax temperature, and MT for node-wise routing over fixed-temperature
experts. Branch outputs are added to the preserved baseline path through
learnable scalar gates (Eq.~\ref{eq:residual}). In the scoring diagrams,
bidirectional arrows denote symmetric scoring ($e_{ij}=e_{ji}$), whereas
directed arrows denote asymmetric scoring ($e_{ij}\neq e_{ji}$), rather than
changes to graph connectivity. The figure uses matrix-level notation for the
layer input and output, while the equations use the corresponding node-level
notation.}
\label{fig:architecture}
\end{figure}

\subsection{Controlled LoRA-AASIST2 Setting}
\label{sec:controlled_setting}
We study graph-attention design choices in AASIST2 under a controlled
LoRA-adapted anti-spoofing pipeline. The base system uses an XLS-R-300M
SSL encoder followed by the AASIST2 backend. Our intervention is restricted
to the spectral and temporal graph-attention layers of AASIST2
(\texttt{gat\_s} and \texttt{gat\_t} in the implementation). Across all
variants, we keep the SSL backbone, LoRA recipe, pooling module, classifier,
and training protocol unchanged. This controlled setting reduces confounding
from the front-end, adaptation strategy, and training configuration, making
the graph-attention interventions directly comparable.
Figure~\ref{fig:architecture} summarizes the controlled intervention scope:
the XLS-R encoder and non-GAT backend components are held fixed, while the
AASIST2 graph-attention layer is augmented with gated residual branches for
the three design axes detailed in Section~\ref{sec:axes} and combined in
Section~\ref{sec:framework}.

\subsection{Baseline AASIST2 Graph Attention}
\label{sec:prelim}
Each targeted AASIST2 graph-attention layer, either spectral or temporal,
maps a node feature matrix
$\mathbf{H}=[\mathbf{h}_1,\ldots,\mathbf{h}_N]^\top$,
$\mathbf{h}_i\in\mathbb{R}^{d}$, to updated node representations.
The baseline attention score between nodes $i$ and $j$ is computed
with an element-wise-product scorer:
\begin{equation}
e_{ij}^{\mathrm{base}} =
\mathbf{a}^{\top}
\tanh\!\left(\mathbf{W}_{p}(\mathbf{h}_i \odot \mathbf{h}_j)\right),
\label{eq:base_score}
\end{equation}
where $\mathbf{W}_{p}$ and $\mathbf{a}$ are learned parameters and
$\odot$ denotes element-wise multiplication. A fixed temperature
$T_0=2.0$ normalizes the scores before neighbor aggregation:
\begin{equation}
\alpha_{ij}^{\mathrm{base}} =
\mathrm{softmax}_{j}\!\left(\frac{e_{ij}^{\mathrm{base}}}{T_0}\right),
\qquad
\tilde{\mathbf{h}}_i^{\mathrm{base}} =
\sum\nolimits_j \alpha_{ij}^{\mathrm{base}}\mathbf{h}_j .
\label{eq:base_att}
\end{equation}
Eq.~\eqref{eq:base_att} abstracts away the output projection,
self-projection, batch normalization, and SELU activation used in
the implementation; we denote the full baseline layer output by
$\mathbf{h}_i^{\mathrm{base}}$. This formulation reveals two inherited
design properties. First, the score is symmetric because
$\mathbf{h}_i\odot\mathbf{h}_j=\mathbf{h}_j\odot\mathbf{h}_i$,
so $e_{ij}^{\mathrm{base}}=e_{ji}^{\mathrm{base}}$ before softmax
normalization. Second, the attention temperature is fixed: its value is
hard-coded to $T_0=2.0$ and shared globally across all nodes. We use these
properties to derive three graph-attention design axes.

\subsection{Graph-Attention Design Axes}
\label{sec:axes}
We decompose the AASIST2 graph-attention design space into three
independently testable axes. The first axis tests scoring symmetry, the
second tests temperature learnability, and the third tests routing granularity.
Each axis is instantiated as a residual branch and later combined through
the gated framework in Section~\ref{sec:framework}.

\textbf{Axis 1: Scoring Symmetry.} The baseline element-wise-product
scoring (Eq.~\ref{eq:base_score}) is symmetric, meaning
$e_{ij} = e_{ji}$. To test whether asymmetric, query-adaptive scoring is
beneficial under LoRA, we add a concat-based scoring
branch~\cite{brody2022iclr_gatv2}:
\begin{equation}
e_{ij}^{\mathrm{gatv2}} =
\mathbf{a}_{g}^{\top}
\mathrm{LeakyReLU}\!\left(\mathbf{W}_{g}(\mathbf{h}_i \| \mathbf{h}_j)\right),
\label{eq:gatv2_score}
\end{equation}
In Eq.~\eqref{eq:gatv2_score}, $\|$ denotes concatenation and
$\mathbf{W}_{g},\mathbf{a}_{g}$
are learned branch-specific parameters. Since
$(\mathbf{h}_i\|\mathbf{h}_j)$ generally differs from
$(\mathbf{h}_j\|\mathbf{h}_i)$, this GATv2-style branch can produce
$e_{ij}^{\mathrm{gatv2}}\neq e_{ji}^{\mathrm{gatv2}}$, breaking the
symmetric scoring constraint of the baseline. The branch uses the same
fixed temperature $T_0$ as the baseline and produces a full layer output
$\mathbf{h}_i^{\mathrm{gatv2}}$.

\textbf{Axis 2: Temperature Learnability.} The baseline fixes $T_0=2.0$
for both targeted graph-attention layers. Inspired by prior use of
temperature scaling in graph attention~\cite{borodin2024asvspoof_aasist3},
we instead make the temperature trainable by adding a branch that replaces
the constant with a learned scalar for each targeted layer:
\begin{equation}
T_{\ell} = \exp(\tau_{\ell}), \quad
\tau_{\ell} \leftarrow \ln T_0 \;\; \text{at initialization}.
\label{eq:learn_t}
\end{equation}
The exponential parameterization in Eq.~\eqref{eq:learn_t} ensures a
positive temperature. This branch uses the same element-wise-product scoring
form as the baseline, with independently initialized scoring and projection
parameters, but replaces the fixed temperature with $T_{\ell}$. At
initialization, the branch uses the same temperature as the baseline; during
training, $T_{\ell}$ can move away from $T_0$ when a different temperature
is beneficial. The resulting full layer output is denoted
$\mathbf{h}_i^{\mathrm{lt}}$.

\textbf{Axis 3: Routing Granularity.} The baseline shares a single global
temperature across all nodes. To test whether different nodes benefit from
different fixed temperatures, we add a multi-expert branch with per-node
routing~\cite{chen2025arxiv_gnnmoe}:
\begin{equation}
\alpha_{ij}^{(k)} =
\mathrm{softmax}_{j}\!\left(\frac{e_{ij}^{\mathrm{mt}}}{T_k}\right),
\qquad
T_k\in\{0.5,2.0,8.0\},
\label{eq:mt}
\end{equation}
In Eq.~\eqref{eq:mt}, all experts share the same element-wise-product scoring logits
$e_{ij}^{\mathrm{mt}}$ but normalize them with different fixed
temperatures. The $k$-th expert output is
$\tilde{\mathbf{h}}_i^{(k)}=\sum_j\alpha_{ij}^{(k)}\mathbf{h}_j$.
A node-wise router then mixes the expert outputs:
\begin{equation}
\mathbf{g}_i =
\mathrm{softmax}\!\left(\frac{\mathrm{MLP}(\mathbf{h}_i)}{\rho}\right),
\qquad
\tilde{\mathbf{h}}_i^{\mathrm{mt}} =
\sum_{k=1}^{K} g_{ik}\tilde{\mathbf{h}}_i^{(k)} ,
\label{eq:mt_router}
\end{equation}
In Eq.~\eqref{eq:mt_router}, $\mathbf{g}_i\in\mathbb{R}^{K}$ and
$\rho$ is a learnable router temperature initialized to 1.0.
This routing is applied to the
aggregated expert representations rather than to the raw attention scores.
The resulting full layer output is denoted $\mathbf{h}_i^{\mathrm{mt}}$.

\subsection{Gated Residual Branch Framework}
\label{sec:framework}
We test the three axes through a common residual interface rather than by
directly replacing the baseline layer. This preserves the original AASIST2
attention path in every variant and provides an architecture-level
degeneration path when a residual gate approaches zero. The output of each
graph-attention layer is
\begin{equation}
\mathbf{h}_i^{\mathrm{out}} =
\mathbf{h}_i^{\mathrm{base}}
+ \lambda_1\mathbf{h}_i^{\mathrm{gatv2}}
+ \lambda_2\mathbf{h}_i^{\mathrm{lt}}
+ \lambda_3\mathbf{h}_i^{\mathrm{mt}},
\label{eq:residual}
\end{equation}
where $\lambda_1$, $\lambda_2$, and $\lambda_3$ are learnable scalar gates
for the GATv2, LearnT, and MT branches, respectively, and are initialized to
a small positive value. A disabled branch is omitted from Eq.~\eqref{eq:residual}.
This framework supports controlled single-axis comparisons and branch
composition tests.

We define six branch configurations including the original baseline
(Table~\ref{tab:configs}). The GATv2 name denotes the GATv2-style
concat-scoring branch used to test the scoring-symmetry axis.

\begin{table}[t]
\centering
\caption{Branch configurations used in the controlled study. The Baseline
uses the original AASIST2 graph-attention layer without residual branches.
Each non-baseline configuration isolates one design axis or tests a
two-branch combination under the gated residual framework
(Eq.~\ref{eq:residual}). $\checkmark$ = branch enabled; $-$ = disabled.}
\label{tab:configs}
\begin{tabular*}{\columnwidth}{@{\extracolsep{\fill}}cccc@{}}
\toprule
Variant & GATv2 ($\lambda_1$) & LearnT ($\lambda_2$) & MT ($\lambda_3$) \\
\midrule
Baseline      & $-$ & $-$ & $-$ \\
GATv2         & $\checkmark$ & $-$ & $-$ \\
LearnT        & $-$ & $\checkmark$ & $-$ \\
MT            & $-$ & $-$ & $\checkmark$ \\
GATv2+LearnT  & $\checkmark$ & $\checkmark$ & $-$ \\
GATv2+MT      & $\checkmark$ & $-$ & $\checkmark$ \\
\bottomrule
\end{tabular*}
\end{table}

In addition to these configurations, we evaluate a FixedT-Res control with
the same residual element-wise-product branch structure as LearnT but with
temperature fixed at $T_0=2.0$. This control separates the effect of adding
residual branch capacity from making the temperature trainable. We omit the
MT+LearnT and GATv2+MT+LearnT combinations because MT and LearnT both
manipulate temperature but through distinct mechanisms (fixed multi-temperature
routing vs.\ a single learned scalar), and combining them would conflate two
temperature effects within a single variant.

\section{Experiments}
\label{sec:experiments}
\subsection{Datasets}
\label{sec:datasets}
We evaluate on one in-domain and four out-of-domain test sets that
differ from the training distribution along distinct axes.
Table~\ref{tab:datasets} summarizes the datasets used for training,
validation, and evaluation.

\begin{table}[t]
\centering
\caption{Summary of datasets used in the experiments. The horizontal
line separates training and validation from evaluation.}
\label{tab:datasets}
\resizebox{\columnwidth}{!}{%
\begin{tabular}{l l c c}
\toprule
Dataset & Usage & \#Bonafide & \#Spoofed \\
\midrule
ASVspoof 2019 LA Train~\cite{wang2020csl_asvspoof2019} & Training & 2,580 & 22,800 \\
ASVspoof 2019 LA Dev~\cite{wang2020csl_asvspoof2019} & Validation & 2,548 & 22,296 \\
\midrule
ASVspoof 2019 LA Eval~\cite{wang2020csl_asvspoof2019} & In-domain & 7,355 & 63,882 \\
ASVspoof 2021 LA~\cite{yamagishi2021asvspoof_asvspoof2021} & Out-of-domain & 18,452 & 163,114 \\
ASVspoof 2021 DF~\cite{yamagishi2021asvspoof_asvspoof2021} & Out-of-domain & 22,617 & 589,212 \\
WaveFake~\cite{frank2021neurips_wavefake} & Out-of-domain & 13,100 & 91,700 \\
In-the-Wild~\cite{muller2022interspeech_itw} & Out-of-domain & 19,963 & 11,816 \\
\bottomrule
\end{tabular}%
}
\end{table}
All models are trained on the Logical Access (LA) partition of
ASVspoof~2019~\cite{todisco2019interspeech_asvspoof,wang2020csl_asvspoof2019},
with the development set used for model selection.
We evaluate on ASVspoof~2019~LA Eval~\cite{wang2020csl_asvspoof2019} as
the in-domain test set and on four out-of-domain sets: ASVspoof~2021~LA
and DF~\cite{yamagishi2021asvspoof_asvspoof2021},
WaveFake~\cite{frank2021neurips_wavefake}, and
In-the-Wild~\cite{muller2022interspeech_itw}.
Together, these evaluation sets cover unseen attack types, codec and channel
variation, neural-vocoder generation, and real-world public-media speech
deepfakes.

\subsection{Evaluation Metrics}
\label{sec:metrics}
We use a unified score-based protocol for all test sets. The primary metric
is Equal Error Rate (EER), defined as the error rate at the operating point
where the false acceptance and false rejection rates are equal. We call this
metric Unified EER because the same score format and threshold-sweeping
procedure are used for ASVspoof, WaveFake, and In-the-Wild, rather than using
dataset-specific official scorers. For each seed, the Avg column is the
unweighted mean of the five dataset-level Unified EER values; the reported
mean and standard deviation are then computed over the five seeds. As
auxiliary metrics, we report Area Under the ROC Curve (AUROC) and Average
Precision (AP).

\subsection{Training Configuration}
\label{sec:training}
All LoRA-scale configurations share the same SSL front-end, LoRA setup,
pooling module, classifier, optimizer, and training protocol; only the
graph-attention intervention differs. The original XLS-R-300M pretrained
backbone weights are kept frozen, while LoRA adapters and LayerNorm
parameters are trainable. The LoRA adapters~\cite{hu2022iclr_lora} with rank $r = 16$, scaling factor
$\alpha = 16$, and dropout $0.05$ are inserted into the query and value
projections ($q_{\text{proj}}$, $v_{\text{proj}}$) of all 24~transformer
layers. Across variants,
the number of trainable parameters ranges from 2.51M to 2.57M,
corresponding to approximately $0.79$--$0.81\%$ of the full model. The
learnable branch gates $\lambda_1, \lambda_2, \lambda_3$ are initialized
to $0.05$.

All reported models are trained with cross-entropy over additive-margin
softmax (AM-Softmax) logits. Given feature vector $\mathbf{f}$ and class
weight $\mathbf{w}_c$, the classifier uses normalized cosine logits
$z_c=s\cos(\mathbf{f},\mathbf{w}_c)$ for non-target classes and
$z_y=s(\cos(\mathbf{f},\mathbf{w}_y)-m)$ for the target class, with
scale $s=15$ and margin $m=0.3$. Label smoothing, center loss, and
feature-level mixup are disabled in the reported XLS-R experiments.

We use AdamW~\cite{loshchilov2019iclr_adamw} with learning rate $10^{-4}$
and weight decay $10^{-4}$. A warmup-cosine schedule is applied with $5\%$
warmup and a minimum learning rate of $5\%$ of the peak value. All models
are trained for 30~epochs with batch size~16, single-precision (FP32)
training, and 64,600-sample inputs (approximately 4~seconds at 16~kHz).
Validation is performed every 2~epochs on ASVspoof 2019 LA Dev, and the
checkpoint used for test evaluation is selected by the lowest
development-set EER. Every configuration is trained independently under five
random seeds $\{1234, 2345, 3456, 5678, 6789\}$. The seed controls parameter
initialization, data shuffling, dataloader workers, and the main
random number generators, enabling a direct assessment of training
stability.

\subsection{Main Results}
\label{sec:main_results}
\begin{figure}[htbp]
\centering
\includegraphics[width=\columnwidth]{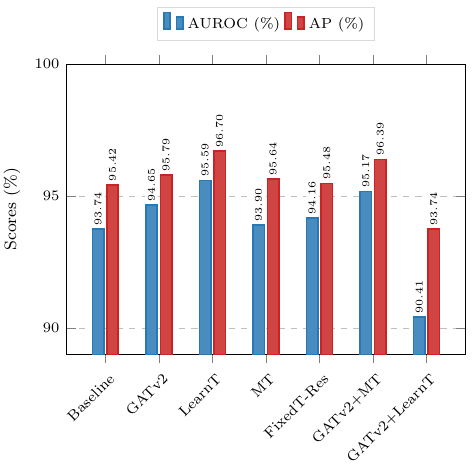}
\caption{Average AUROC and AP across the five evaluation sets. LearnT
achieves the highest AUROC and AP, followed by GATv2+MT, consistent with
the Unified EER ranking in Table~\ref{tab:eer}. Higher is better.}
\label{fig:aux_metrics}
\end{figure}

Table~\ref{tab:eer} reports Unified EER across the five evaluation sets,
and Figure~\ref{fig:aux_metrics} visualizes AUROC and AP averaged over the same sets.
Lower EER is better, whereas higher AUROC and AP are better. In addition
to the controlled variants from Table~\ref{tab:configs}, we include the
FixedT-Res control to separate temperature learnability from residual branch
capacity.

\begin{table*}[t]
\centering
\caption{Unified EER (\%) across five test sets. Mean and standard deviation
are calculated over five random seeds. Best and second-best results among
LoRA-scale configurations are shown in bold and underlined, respectively.
Lower is better.}
\label{tab:eer}
\setlength{\tabcolsep}{0pt}
\begin{tabular*}{\textwidth}{@{\extracolsep{\fill}}ccccccc@{}}
\toprule
Method & ASV19 LA & ASV21 LA & DF21 & WaveFake & In-the-Wild & Avg \\
\midrule
Baseline & 0.63$\pm$0.28 & 7.22$\pm$0.65 & 8.02$\pm$0.76 & 21.24$\pm$3.88 & 16.49$\pm$1.77 & 10.72$\pm$1.24 \\
\midrule
GATv2 & 0.61$\pm$0.21 & 7.22$\pm$0.88 & 7.48$\pm$1.14 & 17.77$\pm$2.30 & \underline{15.04}$\pm$2.33 & 9.62$\pm$1.18 \\
LearnT & 0.63$\pm$0.11 & \underline{6.81}$\pm$0.70 & \textbf{6.58}$\pm$1.03 & \underline{17.06}$\pm$2.22 & \textbf{13.90}$\pm$3.01 & \textbf{8.99}$\pm$1.11 \\
MT & 0.65$\pm$0.26 & 7.75$\pm$1.50 & 9.19$\pm$4.66 & 19.24$\pm$5.49 & 16.73$\pm$6.27 & 10.71$\pm$3.42 \\
FixedT-Res & \textbf{0.55}$\pm$0.11 & 7.54$\pm$1.47 & 8.29$\pm$2.70 & 19.61$\pm$1.89 & 15.18$\pm$4.21 & 10.23$\pm$1.97 \\
GATv2+MT & \underline{0.60}$\pm$0.20 & \textbf{6.77}$\pm$0.11 & \underline{7.05}$\pm$0.27 & \textbf{16.47}$\pm$0.82 & 15.28$\pm$0.78 & \underline{9.23}$\pm$0.19 \\
GATv2+LearnT & 1.36$\pm$1.20 & 7.83$\pm$1.74 & 9.90$\pm$7.24 & 29.37$\pm$12.75 & 18.86$\pm$6.30 & 13.46$\pm$5.19 \\
\bottomrule
\end{tabular*}
\end{table*}

We additionally report official ASVspoof scorer results in
Table~\ref{tab:official_eer}. This evaluation uses the official ASVspoof
scoring protocol and is restricted to the three ASVspoof test sets, since
WaveFake and In-the-Wild do not provide ASVspoof official scoring files.
Therefore, Unified EER over all five test sets remains our primary metric,
while official ASVspoof EER serves as a protocol-specific consistency check
and is not included in the five-set Avg column.

\begin{table}[t]
\centering
\caption{Official ASVspoof EER (\%) computed with the official ASVspoof
scoring protocol and averaged over ASVspoof 2019 LA, ASVspoof 2021 LA, and
ASVspoof 2021 DF. Mean and standard deviation are computed over five random
seeds. Relative reduction is computed against the Baseline; negative values
indicate degradation. Lower EER is better.}
\label{tab:official_eer}
\begin{tabular*}{\columnwidth}{@{\extracolsep{\fill}}ccc@{}}
\toprule
Method & Official Avg & Rel. Reduction \\
\midrule
Baseline & 2.88$\pm$0.36 & -- \\
\midrule
GATv2 & 2.62$\pm$0.27 & $+9.0\%$ \\
LearnT & \textbf{2.38$\pm$0.22} & \textbf{$+17.3\%$} \\
MT & 3.35$\pm$1.81 & $-16.3\%$ \\
FixedT-Res & 2.72$\pm$0.58 & $+5.4\%$ \\
GATv2+MT & \underline{2.46$\pm$0.20} & \underline{$+14.4\%$} \\
GATv2+LearnT & 4.49$\pm$3.58 & $-56.0\%$ \\
\bottomrule
\end{tabular*}
\end{table}

LearnT achieves the best average Unified EER among the LoRA-scale
configurations, reducing the Baseline from $10.72\%$ to $8.99\%$.
GATv2+MT obtains the second-best average EER ($9.23\%$) and the
smallest seed-level standard deviation in the Avg column ($0.19\%$).
By contrast, MT alone is nearly unchanged from the Baseline
($10.71\%$ vs.\ $10.72\%$), and GATv2+LearnT degrades to $13.46\%$.
The auxiliary metrics in Figure~\ref{fig:aux_metrics} follow the same
ranking: LearnT achieves the
highest average AUROC ($95.59\%$) and AP ($96.70\%$), followed by
GATv2+MT ($95.17\%$ AUROC and $96.39\%$ AP). The official ASVspoof results
are consistent with this ranking: LearnT gives the best official average
EER, and GATv2+MT remains the second-best LoRA-scale configuration.

\subsection{Analysis of Model Variants}
\label{sec:main-findings}
Unless noted otherwise, all relative improvements discussed in this section are
computed based on the average Unified EER column of Table~\ref{tab:eer}.
\paragraph{LearnT Gives the Strongest Single-Axis Gain.}
LearnT reduces the average Unified EER from $10.72\%$ to $8.99\%$,
corresponding to a $16.1\%$ relative reduction over the Baseline. This is
the largest mean improvement among the single-axis variants: GATv2 reduces
average EER to $9.62\%$ ($10.2\%$ relative reduction), whereas MT remains
nearly unchanged at $10.71\%$. The FixedT-Res control reaches
$10.23\%\pm1.97$, which improves only modestly over the Baseline and remains
$1.24$ percentage points worse than LearnT. This comparison indicates that
the LearnT gain is not explained by residual branch capacity alone; the
learnable temperature contributes beyond adding another fixed-temperature
branch.

\paragraph{Branch Composition Is Non-Additive.}
The two tested branch combinations behave in opposite ways. GATv2+MT
improves average EER to $9.23\%$ and reduces the Avg standard deviation to
$0.19\%$, even though MT alone provides no average gain and has the largest
single-axis standard deviation ($3.42\%$). In contrast, GATv2+LearnT
combines two individually beneficial branches but increases average EER to
$13.46\%$, a $25.6\%$ relative degradation from the Baseline. The
degradation appears on every evaluation set: ASV19~LA increases from
$0.63\%$ to $1.36\%$, ASV21~LA from $7.22\%$ to $7.83\%$, DF21 from
$8.02\%$ to $9.90\%$, WaveFake from $21.24\%$ to $29.37\%$, and
In-the-Wild from $16.49\%$ to $18.86\%$. These results show that
graph-attention branches do not combine additively under the fixed LoRA
setting.

\paragraph{Improvements Concentrate on Out-of-Domain Sets.}
LearnT leaves the in-domain ASVspoof~2019~LA EER essentially unchanged
($0.63\%$ for both Baseline and LearnT), but improves all four
out-of-domain sets. The out-of-domain average EER decreases from
$13.24\%$ for the Baseline to $11.09\%$ for LearnT, a $16.3\%$ relative
reduction. The strongest absolute reduction occurs on WaveFake, where EER
drops from $21.24\%$ to $17.06\%$. GATv2+MT is less accurate on average
than LearnT, but it is the most stable configuration: its Avg standard
deviation is $0.19\%$ compared with $1.24\%$ for the Baseline and $1.11\%$
for LearnT, and its out-of-domain coefficient of variation is $3.90\%$
compared with $11.88\%$ for the Baseline. We therefore treat GATv2+MT as a
stability-oriented composition rather than the best mean-EER variant.

\section{Discussion and Implications}
\label{sec:discussion}

The results show that the AASIST2 backend remains an important design lever
in LoRA-adapted SSL anti-spoofing systems. Recent systems draw on SSL
encoders, graph-attention backends, and parameter-efficient
adaptation~\cite{wang2022odyssey_ssl_frontends,babu2022interspeech_xlsr,zhang2024icassp_aasist2,hu2022iclr_lora,laakkonen2025arxiv_metalora,xia2024taslp_molex,chen2025amulet},
which makes it difficult to attribute gains to a specific component. By fixing
the encoder, adapter placement, classifier, optimizer, and data protocol, our
experiments isolate graph-attention design under the same adaptation budget.
The resulting spread is substantial: among LoRA-scale variants, the best and
worst average Unified EER differ by $4.47$ percentage points despite nearly
identical trainable parameter counts. Thus, the AASIST2 graph-attention layer
is not merely an implementation detail behind the SSL front-end, but a
meaningful source of cross-domain behavior.

Among the tested axes, the strongest positive result comes from temperature
learnability rather than added branch capacity. LearnT outperforms both the
Baseline and the FixedT-Res control, separating the effect of a trainable
temperature from the effect of adding another residual graph-attention path.
This aligns with work on calibration and optimization
dynamics~\cite{guo2017icml_calibration,hsu2022neurips_gats,agarwala2023temperature}
and extends fixed-temperature anti-spoofing studies toward a trainable
setting~\cite{borodin2024asvspoof_aasist3,firc2026spaarsist}.
The learned temperature remains close to the original AASIST2 value, averaging
$1.96$ after initialization at $T_0=2.0$, which suggests limited adaptive
control over softmax sharpness rather than a different attention regime.

The branch-composition results provide the complementary caution. GATv2 and
LearnT each improve average EER alone, but GATv2+LearnT degrades below the
Baseline on every evaluation set; conversely, MT is weak as a single branch,
yet GATv2+MT gives the second-best average EER and the lowest seed-level
standard deviation. Backend attention should therefore be treated as a coupled
system of scoring, normalization, and routing decisions. Since the AASIST2
graph topology is held fixed, these conclusions concern attention mechanisms
rather than graph construction. Practically, LearnT is preferable when mean
out-of-domain EER is the target, whereas GATv2+MT is preferable when
seed-level stability is prioritized. Evaluations should report both
out-of-domain performance and seed-level stability, in line with
cross-dataset evaluation practice in audio deepfake
detection~\cite{das2020icassp_generalized_cm,yamagishi2021asvspoof_workshop,yamagishi2021asvspoof_asvspoof2021,muller2021asvspoof_silence,muller2022interspeech_itw,muller2024harder,kwok2025interspeech_bonafide_cross,wang2026csl_asvspoof5}.

\section{Conclusion}
\label{sec:conclusion}
We presented a controlled study of graph-attention design choices in a
LoRA-adapted XLS-R-300M + AASIST2 anti-spoofing system. By fixing the SSL
front-end, LoRA configuration, classifier, and training protocol, we isolated
three backend axes: scoring symmetry, temperature learnability, and routing
granularity. The results show that graph-attention design remains important
under a strict LoRA-scale parameter budget: LearnT achieves the best average
Unified EER and reduces out-of-domain EER from $13.24\%$ to $11.09\%$,
whereas GATv2+MT provides the lowest seed-level standard deviation. This
out-of-domain gain indicates improved cross-domain generalization rather than
only better in-domain fitting. However, the degradation of GATv2+LearnT shows
that backend interventions should be evaluated as interacting components
rather than isolated plug-ins.Future work will extend this controlled analysis to other SSL encoders, adapter settings, and graph-based backends.

\bibliographystyle{IEEEtran}
\bibliography{reference}

\end{document}